# On Distance Properties of Quasi-Cyclic Protograph-Based LDPC Codes

Brian K. Butler and Paul H. Siegel
Department of Electrical & Computer Engineering, University of California San Diego, La Jolla, CA, USA
{bkbutler, psiegel}@ucsd.edu

***Abstract*—Recent work [1][2] has shown that properly designed protograph-based LDPC codes may have minimum distance linearly increasing with block length. This notion rests on ensemble arguments over all possible expansions of the base protograph. When implementation complexity is considered, the expansion is typically chosen to be quite orderly. For example, protograph expansion by cyclically shifting connections creates a quasi-cyclic (QC) code. Other recent work [3] has provided upper bounds on the minimum distance of QC codes. In this paper, these bounds are expanded upon to cover puncturing and tightened in several specific cases. We then evaluate our upper bounds for the most prominent protograph code thus far, one proposed for deep-space usage in the CCSDS experimental standard [4], the code known as AR4JA.**

## I. Introduction

An important class of modern codes, the Low Density Parity Check (LDPC) codes, had their start in the seminal work by R. Gallager [5] in 1963. Properly designed, LDPC codes exhibit very low SNR thresholds in their error rate performance. However there has been a tradeoff evident between SNR threshold and error floor performance. An early technique to lower error floors in LDPC was to reduce the number of small cycles in the graph and "neighborhood" optimize short loop multiplicities [6]. Similarly, the ACE algorithm [7] for placing edges in a graph-based code brings down the error floor substantially by preventing small cycles from clustering on low degree variables.

However, one important component limiting the error floor of any code is the minimum distance, and work on designing LDPC codes for large minimum distance has been limited. The minimum distance is also important in understanding the likelihood of undetectable error patterns which are critical to limit in certain applications such as data storage.

Code ensembles based on protographs with certain properties have been shown to achieve a minimum distance linearly increasing with block length [1][2] – a powerful feature for floor performance. These protographs together with the ACE algorithm have been used to design LDPC codes for deep-space usage in the CCSDS experimental standard [4]. The standard's codes, as specified, also fall into the class of Quasi-Cyclic (QC) codes. A separate body of work on QC LDPC codes exists, including recent work on distance upper bounds [3]. We attempt to bring these works together by extending the bounds to punctured LDPC codes and tightening the bounds where possible.

## II. Protographs & AR4JA

Protographs were introduced as a way to impart structure to the inter-connectivity of graph-based codes [8]. Protographs themselves are a subset of the multi-edge type graphs introduced in [9].

A *protograph* is a Tanner graph with a relatively small number of nodes, except parallel edges are permitted. A protograph, $G = (V,C,E)$, consists of a set of variable nodes $V$, a set of check nodes $C$, and a set of edges $E$. Each edge, $e \in E$, connects a variable node, $v_e \in V$, to a check node, $c_e \in C$. A useful refinement is to allow the variable node set $V$ to contain untransmitted or punctured variables.

This work was supported in part by the Center for Magnetic Recording Research at the University of California, San Diego (UCSD) and by the National Science Foundation (NSF) under Grant CCF-0829865.

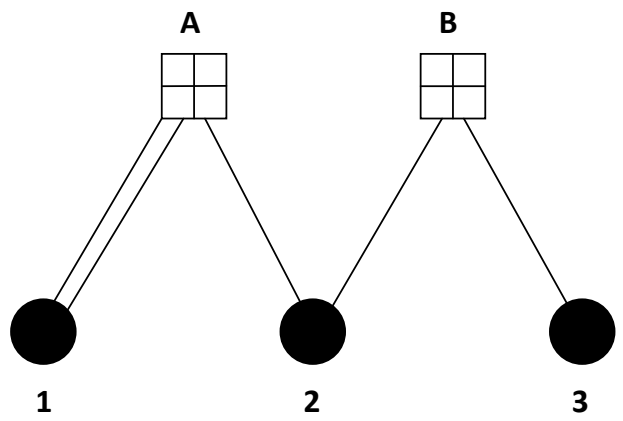

$$\mathbf{A} = \begin{bmatrix} 2 & 1 & 0 \\ 0 & 1 & 1 \end{bmatrix}$$

Figure 1. Simple protograph and corresponding protomatrix.

A simple protograph is shown in Fig. 1 with three variable nodes, two check nodes, and five edges. The accompanying protomatrix in Fig. 1 fully describes the graph. The labeling of the protograph indicates node *types*. All copies of check node A, are termed "type A" check nodes. Similarly, all copies of variable node 1, are termed "type 1" variable nodes.

The *derived graph* is created by replicating the protograph many times and interconnecting the copies. The *protograph code* is defined by the resulting derived graph. The interconnection process proceeds by treating each set of edge copies as an edge set, and swapping connections only within each edge set. This rule prevents nodes changing degree and maintains the graph's connectivity by node type.

The main advantages of protographs are that degree one variable nodes and untransmitted ("punctured") variable nodes may be introduced in a structured way. Interestingly, the optimization of irregular LDPC codes by density evolution does not allow for degree one variable nodes, but produces a significant number of degree two nodes. An additional advantage of protographs is that decoder hardware should be less complex due to the local structure.

The parity check matrix corresponding to a possible derived graph after making $N = 3$ copies of the protograph of Fig. 1 is shown below. When divided into 3×3 submatrices, the correspondence back to the protomatrix, **A**, of Fig. 1 is evident.

$$\mathbf{H} = \begin{pmatrix} 1 & 0 & 1 & 1 & 0 & 0 & 0 & 0 & 0 \\ 1 & 1 & 0 & 0 & 1 & 0 & 0 & 0 & 0 \\ 0 & 1 & 1 & 0 & 0 & 1 & 0 & 0 & 0 \\ 0 & 0 & 0 & 0 & 1 & 0 & 1 & 0 & 0 \\ 0 & 0 & 0 & 0 & 0 & 1 & 0 & 1 & 0 \\ 0 & 0 & 0 & 1 & 0 & 0 & 0 & 0 & 1 \end{pmatrix}$$

The AR4JA protograph [1] for code rate ½ is shown in Fig. 2, following the convention of showing transmitted variables as solid circles and the untransmitted variable as an outlined circle. The protograph of the rate-½ code is extended to rate-2/3 by adding two degree-four variable nodes. The corresponding protomatrices are shown in (1) and (2), respectively. The variables have been numbered in the figure to correspond to columns of the protomatrix from left-to-right. The AR4JA family of protographs continues to increase the offered code rate options by adding more pairs of degree-four variables connected to just two of the check nodes. In all cases, it is the variables corresponding to the right-most column of the protomatrix that are punctured (not transmitted), i.e., those variables of degree six.

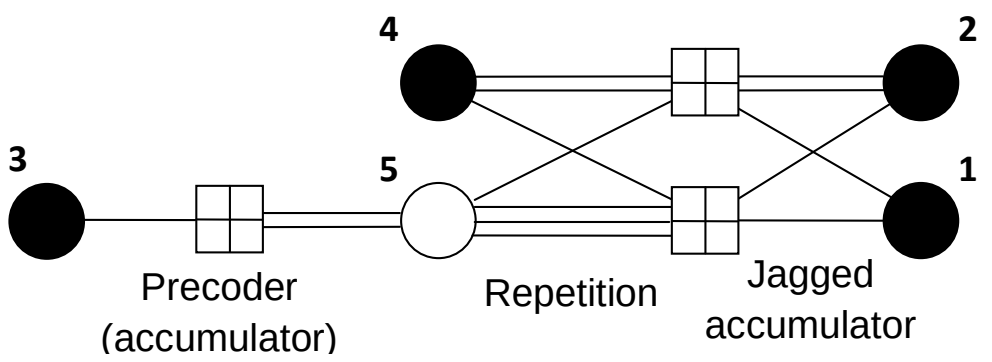

Figure 2. AR4JA Protograph, rate ½. The transmitted variables are shown as solid circles, the untransmitted variable as an outlined circle.

$$\mathbf{A}_{r=1/2} = \begin{bmatrix} 0 & 0 & 1 & 0 & 2 \\ 1 & 1 & 0 & 1 & 3 \\ 1 & 2 & 0 & 2 & 1 \end{bmatrix} \tag{1}$$

$$\mathbf{A}_{r=2/3} = \begin{bmatrix} 0 & 0 & 0 & 0 & 1 & 0 & 2 \\ 3 & 1 & 1 & 1 & 0 & 1 & 3 \\ 1 & 3 & 1 & 2 & 0 & 2 & 1 \end{bmatrix} \tag{2}$$

Techniques for calculating the asymptotic ensemble weight enumerators for protograph-based codes have been presented [1][2][10]. From the derived expression of the weight spectrum, the *typical minimum distance ratio*, $\delta_{min}$, can be found, if it exists. With high probability the minimum distance of most codes in the ensemble increases linearly with n with proportionality constant $\delta_{min}$. The conditions for a protograph-based LDPC code to meet $\delta_{min}$, > 0 are presented in [2]. The AR4JA rate-½ protomatrix of (1) is found to have $\delta_{min}$, = 0.015 [1][2].

## III. QC Expansion

A *quasi-cyclic* (QC) code is a linear block code having the property that applying identical circular shifts to every length-$N$ subblock of a codeword yields a codeword. If there is just a single subblock, the code is also a classic cyclic block code.

A QC LDPC code of length $n = L\,N$ can be described by a $m \times n$ scalar parity-check matrix, $\mathbf{H} \in \mathbb{F}_2^{m \times n}$, with $m = J\,N$. The code can also be described in polynomial form, since there exists an isomorphism between the ring of $N \times N$ circulant matrices and the ring of polynomials of degree less than $N$, $\mathbb{F}_2[x]/\left\langle x^N - 1 \right\rangle$. Addition and multiplication of the polynomials in the ring happens in the usual way, modulo $x^N$-1. (All the rings in this work are commutative rings containing a multiplicative identity element.) A right *circulant matrix* is a square matrix with each successive row right-shifted circularly one position relative to the row above. Hence, circulant matrices can be completely described by a single row or column. We will use the mapping convention of taking the matrix's first column terms top to bottom as polynomial coefficients of increasing order [3]. A polynomial of 1 must correspond to the $N \times N$ identity matrix, which is right circulant. A few examples of the isomorphism for $N = 3$ are shown below.

$$1 \sim \begin{bmatrix} 1 & 0 & 0 \\ 0 & 1 & 0 \\ 0 & 0 & 1 \end{bmatrix} \quad x \sim \begin{bmatrix} 0 & 0 & 1 \\ 1 & 0 & 0 \\ 0 & 1 & 0 \end{bmatrix} \quad 1 + x^2 \sim \begin{bmatrix} 1 & 1 & 0 \\ 0 & 1 & 1 \\ 1 & 0 & 1 \end{bmatrix}$$

This isomorphism requires that care be taken when representing multiplication of a circulant matrix, $\mathbf{M}$, by a vector, $\mathbf{v} = (v_0, v_1, \ldots v_{N-1})$. We can associate the polynomial, $M(x)$, with the matrix using the technique just described, and associate $v(x) = v_0+v_1x+\ldots v_{N-1}x^{N-1}$ with the vector. The representation of multiplying the circulant matrix from the right with the vector is defined simply as $M(x)\,v(x) \bmod (x^N-1)$. While not used in this paper, please note that multiplying the circulant matrix from the left with the vector in this notation, must be represented by polynomials as $x^N M(x^{-1})\,v(x) \bmod (x^N-1)$.

A *permutation matrix* (not necessarily circulant) is a square matrix of ones and zeros, such that sum of each row and each column is one. A *cyclic permutation matrix* is both a permutation matrix and a circulant matrix, described above. As cyclic permutation matrices in the ring $\mathbb{F}_2^{N \times N}$ are isomorphic to monomials in the ring $\mathbb{F}_2[x]/\left\langle x^N - 1 \right\rangle$, they are both units (invertible elements) in their respective rings. Other elements are not necessarily invertible.

As we are interested in the connection between protographs and QC LDPC codes, we focus on parity-check matrices, $\mathbf{H}$, that are in $J \times L$ block matrix form, described by circulant sub-matrices, each $N \times N$. Let

$$\mathbf{H} = \begin{bmatrix} \mathbf{H}_{0,0} & \cdots & \mathbf{H}_{0,L-1} \\ \vdots & \ddots & \vdots \\ \mathbf{H}_{J-1,0} & \cdots & \mathbf{H}_{J-1,L-1} \end{bmatrix},$$

where each submatrix, $\mathbf{H}_{j,i} \in \mathbb{F}_2^{N\times N}$, is circulant and $\mathbf{H}_{j,i} = \sum_{s=0}^{N-1} h_{j,i,s,0}\,\mathbf{I}_s$, where $\mathbf{I}_s$ is the identity matrix circularly left-shifted by $s$ positions. Now, we can write the *polynomial parity-check matrix,* $\mathbf{H}(x) \in \left[\mathbb{F}_2[x]/\left\langle x^N - 1\right\rangle\right]^{J\times L}$,

$$\mathbf{H}(x) = \begin{bmatrix} h_{0,0}(x) & \cdots & h_{0,L-1}(x) \\ \vdots & \ddots & \vdots \\ h_{J-1,0}(x) & \cdots & h_{J-1,L-1}(x) \end{bmatrix},$$

where $h_{j,i}(x) \triangleq \sum_{s=0}^{N-1} h_{j,i,s,0}\, x^s$ .

Further, we will be interested in the weight of each polynomial of $\mathbf{H}(x)$ (or equivalently, submatrix of $\mathbf{H}$). We'll start by defining the weight of each polynomial, $\mathrm{wt}\left(h_{j,i}(x)\right)$, as the number of non-zero coefficients in $h_{j,i}(x)$. Now we're ready to define the $J \times L$ *weight matrix* as,

$$\mathrm{wt}\left(\mathbf{H}(x)\right) \triangleq \begin{bmatrix} \mathrm{wt}\left(h_{0,0}(x)\right) & \cdots & \mathrm{wt}\left(h_{0,L-1}(x)\right) \\ \vdots & \ddots & \vdots \\ \mathrm{wt}\left(h_{J-1,0}(x)\right) & \cdots & \mathrm{wt}\left(h_{J-1,L-1}(x)\right) \end{bmatrix}.$$

A connection back to protographs can be seen here as the resulting QC LDPC weight matrix, above, corresponds directly to the protomatrix of a protograph — one that has been expanded with circulant matrices.

Just as the matrices used to describe QC codes are convenient in polynomial form, so are the resulting vectors. Let polynomial $c(x) \in \mathbb{F}_2[x]/\left\langle x^N - 1\right\rangle$ have weight, $\mathrm{wt}\left(c(x)\right)$, equal to the number of non-zero coefficients. Define a vector of length-$L$ polynomials $\mathbf{c}(x) \in \left[\mathbb{F}_2[x]/\left\langle x^N - 1\right\rangle\right]^L$ to be,

$$\mathbf{c}(x) = \left(c_0(x), c_1(x), \ldots, c_{L-1}(x)\right).$$

In an error-correcting code context one will note that the equivalent condition of $\mathbf{H}\,\mathbf{c}^T = \mathbf{0}^T$ (with elements in $\mathbb{F}_2$) is $\mathbf{H}(x)\mathbf{c}^T(x) = \mathbf{0}^T$ (with elements in $\mathbb{F}_2[x]/\left\langle x^N - 1\right\rangle$).

## IV. QC Expansion Min. Distance Bounds

In this section we extend the Hamming distance upper bounds of [3] to punctured versions of quasi-cyclic LDPC codes.

We will use the shorthand notation of $[L]$, to indicate the set of $L$ consecutive integers, $\{0,1,2,\ldots,L-1\}$. We will use the common backslash notation to exclude a member from a set. For example, the set $\mathcal{S} \setminus i$ contains all the elements of $\mathcal{S}$ except element $i$. Additionally, a subscript will appear on matrices to indicate a submatrix of just the indicated columns — so that $\mathbf{A}_{\mathcal{S}}$ is a submatrix of $\mathbf{A}$ containing just the columns in the set $\mathcal{S}$.

We use the permanent operation on square matrices throughout the remainder of this paper, denoted, perm(**B**). The permanent is similar to the determinant of linear algebra, but without the (±1) multiplicative term. The permanent of a $J \times J$ matrix, $\mathbf{B} = [b_{j,i}]$, is defined to be

$$\text{perm}(\mathbf{B}) \triangleq \sum_{\sigma} \prod_{j \in [J]} b_{j,\sigma(j)} \tag{3}$$

where the summation is over all $J!$ permutations of the set $[J]$. The function $\sigma(j)$ denotes a permutation of the set $[J]$. When the field is of characteristic two, perm(**B**) = det(**B**) as addition and subtraction are interchangeable in GF($2^m$).

We are interested only in puncturing patterns that maintain the quasi-cyclic property and preserve the dimensionality of the code (*i.e.*, information bits per block). By puncturing whole columns of the polynomial parity check matrix, $\mathbf{H}(x)$, we maintain the quasi-cyclic property. Care must be taken throughout this work that puncturing does not reduce the dimensionality of the code.

We begin with an un-punctured code, $\mathcal{C}$, based upon $\mathbf{H}(x)$. Next, we define a new code, $\mathcal{C}'$, by designating a set of columns, denoted $\mathcal{P}$, of $\mathbf{H}(x)$ to be punctured. The columns in $\mathcal{P}$, are a subset of the $L$ columns, $\mathcal{P} \subset [L]$, and of size such that some redundancy remains intact, $|\mathcal{P}| < J$.

**Lemma 1.** *Let* $\mathcal{C}'$ *be a punctured QC code created by puncturing variables of code* $\mathcal{C}$, *which is defined by the polynomial parity check matrix* $\mathbf{H}(x) \in \left[\mathbb{F}_2[x]/\left\langle x^N - 1\right\rangle\right]^{J \times L}$. *The variables in code* $\mathcal{C}$ *corresponding to columns of* $\mathbf{H}(x)$ *contained in set* $\mathcal{P}$, $\mathcal{P} \subset [L]$, *are punctured. Let* $\mathcal{S}$ *be an arbitrary size-*$(J+1)$ *subset of* $[L]$. *Let the length-L vector,* $\mathbf{c}'(x) = \left(c'_0(x), c'_1(x), \ldots, c'_{L-1}(x)\right)$, *with* $c'(x) \in \left\{\mathbb{F}_2[x]/\left\langle x^N - 1\right\rangle, \varnothing\right\}$ *be defined by:*

$$c'_i(x) \triangleq \begin{cases} \text{perm}\left(\mathbf{H}_{\mathcal{S}\setminus i}(x)\right) & \text{if } i \in \mathcal{S} \setminus \mathcal{P} \\ \varnothing \text{ (untransmitted)} & \text{if } i \in \mathcal{P} \\ 0 & \text{otherwise.} \end{cases} \tag{4}$$

*Then* $\mathbf{c}'(x)$ *is codeword of the punctured code* $\mathcal{C}'$.

*Proof:* This follows from keeping $\mathbf{H}(x)$ unchanged between $\mathcal{C}$ and $\mathcal{C}'$, applying Lemma 6 of [3] for the un-punctured code $\mathcal{C}$, and following our choice of the puncturing pattern to create $\mathcal{C}'$. ■

**Theorem 1.** *Let* $\mathcal{C}'$ *be a punctured QC code created by puncturing variables of code* $\mathcal{C}$ *with polynomial parity-check matrix* $\mathbf{H}(x) \in \left[\mathbb{F}_2[x]/\left\langle x^N - 1\right\rangle\right]^{J \times L}$ *while maintaining the dimensionality and let* $\mathbf{A} \triangleq \text{wt}\left(\mathbf{H}(x)\right)$. *Let the set* $\mathcal{P}$, $\mathcal{P} \subset [L]$, *specify the columns of* ***H****(x) that correspond to the punctured variables. Then the minimum Hamming distance of* $\mathcal{C}'$ *can be upper bounded as follows*

$$d_{\min}(\mathcal{C}') \le \min_{\substack{\mathcal{S}\subseteq[L]\\ |\mathcal{S}|=J+1}}^{*} \sum_{i\in\mathcal{S}\setminus\mathcal{P}} \operatorname{perm}\left(\mathbf{A}_{\mathcal{S}\setminus i}\right) \tag{5}$$

*Proof:* Our proof is lengthy and largely parallels the proofs of Theorems 7 & 8 of [3], while accounting for puncturing. ■

## V. New Tighter Bounds On Minimum Distance

Examining the rate-2/3 AR4JA protomatrix (2), we see cases where the selection of four columns of the weight matrix $\mathbf{A}$ will produce an $\mathbf{A}_{\mathcal{S}}$ matrix containing an all-zero row on top. This particular selection of $\mathcal{S}$ produces the all-zero codeword by the codeword construction of Lemma 1. The contributions of this specific $\mathcal{S}$ set to the upper bound of Theorem 1 will be nil. We can improve those bounds by finding non-zero codewords after row elimination.

**Lemma 2.** *Let $\mathcal{C}'$ be a punctured QC code created by puncturing variables of code $\mathcal{C}$, defined by the polynomial parity check matrix $\mathbf{H}(x) \in \left[\mathbb{F}_2[x]/\left\langle x^N - 1\right\rangle\right]^{J\times L}$. The variables in code $\mathcal{C}$ corresponding to columns of $\mathbf{H}(x)$ contained in set $\mathcal{P}$, $\mathcal{P} \subset [L]$, are punctured. Let $\mathbf{H}'(x)$ be a submatrix of $\mathbf{H}(x)$ with rows $\mathbf{h}_{j^*}(x)$, $j^* \in \mathcal{T} \subset [J]$, removed. Let $\mathcal{S}$ be a subset of $[L]$ of size $J+1-|\mathcal{T}|$, such that*

$$f_{\mathbf{H}'}(\mathcal{S}, j^*) \triangleq \operatorname{perm}\left(\begin{bmatrix}\mathbf{h}_{j^*}(x)\\ \mathbf{H}'_{\mathcal{S}}(x)\end{bmatrix}\right) = 0 \;\; \forall \;\; j^* \in \mathcal{T}.$$

*Let the length-L vector, $\mathbf{c}'(x) = \left(c'_0(x), c'_1(x), \ldots, c'_{L-1}(x)\right)$, with $c'(x) \in \left\{\mathbb{F}_2[x]/\left\langle x^N - 1\right\rangle, \varnothing\right\}$ be defined by:*

$$c'_i(x) \triangleq \begin{cases} \operatorname{perm}\left(\mathbf{H}'_{\mathcal{S}\setminus i}(x)\right) & \text{if } i \in \mathcal{S}\setminus\mathcal{P} \\ \varnothing \text{ (untransmitted)} & \text{if } i \in \mathcal{P} \\ 0 & \text{otherwise}. \end{cases} \tag{6}$$

*Then $\mathbf{c}'(x)$ is codeword of the punctured code, $\mathcal{C}'$.*

*Proof:* We break the proof into two parts.

Case 1: If $\mathcal{P} = \{\varnothing\}$ (the code is un-punctured), it can be shown that the vector $\mathbf{c}'(x)$ multiplied by the several pieces of the parity check matrix yields zeros and therefore the vector $\mathbf{c}'(x)$ must be a codeword in the code.

Case 2: If the code is punctured, this lemma follows from keeping $\mathbf{H}(x)$ unchanged, applying this lemma for the un-punctured case, and then following our choice of puncturing pattern. ■

Not only does Lemma 2 help remove single all-zero rows, it helps us produce lower weight codewords in a case such as below.

$$\mathbf{H}_{\mathcal{S}}(x) = \begin{bmatrix} 0 & 0 & 0 \\ 0 & 0 & x^a + x^b \\ x^c & x^d & x^e \end{bmatrix} \quad (7)$$

First, performing single row removal on (7) (noting that the $\text{perm}\left(\mathbf{H}_{\mathcal{S}}(x)\right) = 0$ as required for $|\mathcal{T}|=1$) generates the all-zero codeword or the codeword segment

$$\mathbf{c}_{\mathcal{S}}(x) = \left(x^{d+a} + x^{d+b}, x^{c+a} + x^{c+b}, 0\right) \bmod \left\langle x^N - 1 \right\rangle.$$

However, looking deeper, Lemma 2 will let us delete two specific rows, $\mathcal{T} = \{0,1\}$, when the column set is $\mathcal{S} = \{0,1\}$ producing the obvious codeword segment $\mathbf{c}_{\mathcal{S}}(x) = (x^d, x^c, 0)$.

**Theorem 2.** *Let $\mathcal{C}'$ be a punctured QC code created by puncturing variables of code $\mathcal{C}$ with polynomial parity-check matrix $\mathbf{H}(x) \in \left[\mathbb{F}_2[x] / \left\langle x^N - 1 \right\rangle\right]^{J \times L}$ while maintaining the dimensionality and let $\mathbf{A} \triangleq \text{wt}\left(\mathbf{H}(x)\right)$. Let the set $\mathcal{P}$, $\mathcal{P} \subset [L]$, specify the columns of **H(x)** that correspond to the punctured variables. Let $\mathbf{H}'(x)$ be a submatrix of $\mathbf{H}(x)$ with rows $\mathbf{h}_{j^*}(x)$, $j^* \in \mathcal{T} \subset [J]$, removed. Let $\mathcal{S}$ be a subset of $[L]$ of size $J+1-|\mathcal{T}|$, such that the sub-rows $\mathbf{a}_{j^*,\mathcal{S}} = (0,0,\ldots,0)\ \forall\, j^* \in \mathcal{T}$, and let $\mathbf{A}'$ be a submatrix of $\mathbf{A}$ with rows $\mathbf{a}_{j^*}$ removed. Then the minimum Hamming distance of $\mathcal{C}'$ is upper bounded as follows*

$$d_{\min}(\mathcal{C}') \le \min_{\substack{|\mathcal{S}|+|\mathcal{T}|=J+1 \\ \mathbf{a}_{j^*,\mathcal{S}}=\mathbf{0}\,\forall\, j^* \in \mathcal{T}}}{}^{*} \sum_{i \in \mathcal{S} \setminus \mathcal{P}} \text{perm}\left(\mathbf{A}'_{\mathcal{S} \setminus i}\right) \quad (8)$$

Proof: Omitted. ■

Below is an example of a weight matrix that will show the benefit of Theorem 2.

$$\mathbf{A} = \begin{bmatrix} 0 & 0 & 3 & 0 & 3 \\ 1 & 1 & 0 & 1 & 3 \\ 1 & 2 & 0 & 2 & 1 \end{bmatrix} \quad (9)$$

Using $\mathbf{A}$ above, treating it as un-punctured, Theorem 1 produces a minimum distance upper bound of 30, while Theorem 2 produces distance bound of just 10. The reason is that Theorem 1 produces zero distances several times when $\mathbf{A}_{\mathcal{S}}$ contains an all-zero row and the bound is only computed when the relatively strong contributions of the 3's in the top row are present. Theorem 2 will remove the top row in one of its calculations and reveal some weaker codeword structure.

## VI. EXPANSION OF AR4JA

A direct QC expansion of the AR4JA protograph shown in Fig. 2 will create a QC LDPC code. Applying the AR4JA protomatrices of (1) & (2) to the bounds given by Theorems 1 & 2 leads to computed upper bounds on the minimum Hamming distance of 10 for all rates, independent of block length. As a Hamming distance of 10 is rather limited for the long block lengths desired, a more involved expansion process is of interest.

TABLE I
MINIMUM DISTANCE OF CCSDS AR4JA PROTOMATRICES STAGE 2

| **Code Rate** | **Upper Bound by Theorems 1 & 2**[a] | $\binom{L}{J+1}$ |
|---|---|---|
| $r$ =1/2 | 66 | ~$7.8\times10^{4}$ |
| $r$ =2/3 | 58 | ~$3.7\times10^{8}$ |
| $r$ =4/5 | 56[b] | ~$5.2\times10^{10}$ |

[a]Row removal of up to 2 rows computed.
[b]Computations are not exhaustive due to complexity.

The AR4JA codes defined in the experimental CCSDS standard [4] use a two step expansion process. After a first cyclic expansion by a factor of 4, a new larger type-I weight matrix is obtained as shown in (10) for rate-½. A *type-I* weight matrix is one that contains only ones and zeros – meaning that the associated protograph has no parallel edges.

According to the CCSDS standard, the matrix (10) is expanded in a second step cyclic expansion to create the three block lengths, corresponding to k=1024, 4 096, and 16 384 information bits, QC LDPC code. In this final expansion, the scalar parity check matrix, **H**, is created by replacing each 1 entry of (10) by a cyclic permutation submatrix selected by a variation on the ACE algorithm. These codes are QC with a subblock size equal to the second step expansion factor. In other words, the two-step process is not equivalent to any single step cyclic expansion. With this in mind, the new protomatrices such as (10) should be used to compute the QC distance upper bounds described here for proper application to the CCSDS AR4JA codes. Those results are shown in Table I. Also a measure of the complexity of completely evaluating Theorem 1 is shown in the Table I.

$$\mathbf{A}=\left[\begin{array}{cccc|cccc|cccc|cccc|cccc}
0&0&0&0&0&0&0&0&1&0&0&0&0&0&0&0&1&0&0&1\\
0&0&0&0&0&0&0&0&0&1&0&0&0&0&0&0&1&1&0&0\\
0&0&0&0&0&0&0&0&0&0&1&0&0&0&0&0&0&1&1&0\\
0&0&0&0&0&0&0&0&0&0&0&1&0&0&0&0&0&0&1&1\\
\hline
1&0&0&0&1&0&0&0&0&0&0&0&1&0&0&0&0&1&1&1\\
0&1&0&0&0&1&0&0&0&0&0&0&0&1&0&0&1&0&1&1\\
0&0&1&0&0&0&1&0&0&0&0&0&0&0&1&0&1&1&0&1\\
0&0&0&1&0&0&0&1&0&0&0&0&0&0&0&1&1&1&1&0\\
\hline
1&0&0&0&0&0&1&1&0&0&0&0&1&1&0&0&1&0&0&0\\
0&1&0&0&1&0&0&1&0&0&0&0&0&1&1&0&0&1&0&0\\
0&0&1&0&1&1&0&0&0&0&0&0&0&0&1&1&0&0&1&0\\
0&0&0&1&0&1&1&0&0&0&0&0&1&0&0&1&0&0&0&1
\end{array}\right] \tag{10}$$

## VII. CONCLUSION

This work has extended the distance bounds of [3] to punctured QC LDPC codes (as required to analyze AR4JA). We have also tightened those distance bounds in several cases that are relevant to protomatrices that contain many zeros per row.

Next we evaluated the minimum distance upper bounds for the AR4JA codes as specified in CCSDS's experimental standard for deep space, by using the protomatrices. We've shown that the 2-step expansion approach was critical for achieving reasonably high minimum distance for these codes in QC LDPC form. We've shown that the minimum Hamming distance of the standardized AR4JA codes do not grow linearly in block length as is the case for the ensemble of AR4JA codes[1][2]. In the ensemble AR4JA analyses, the ensemble of all possible expansions of the base protograph was considered and not the limited number of expansions available when limited to cyclic matrices. While, not linear in block length, the minimum distance at rate-½ is likely high enough for practical purposes. The comparison of the presented Hamming distance measures versus the block length, for rate-½ AR4JA, can be summarized in Fig. 3. Also in Fig. 3, we show the smallest results found using search techniques on the completely expanded CCSDS rate ½ codes.

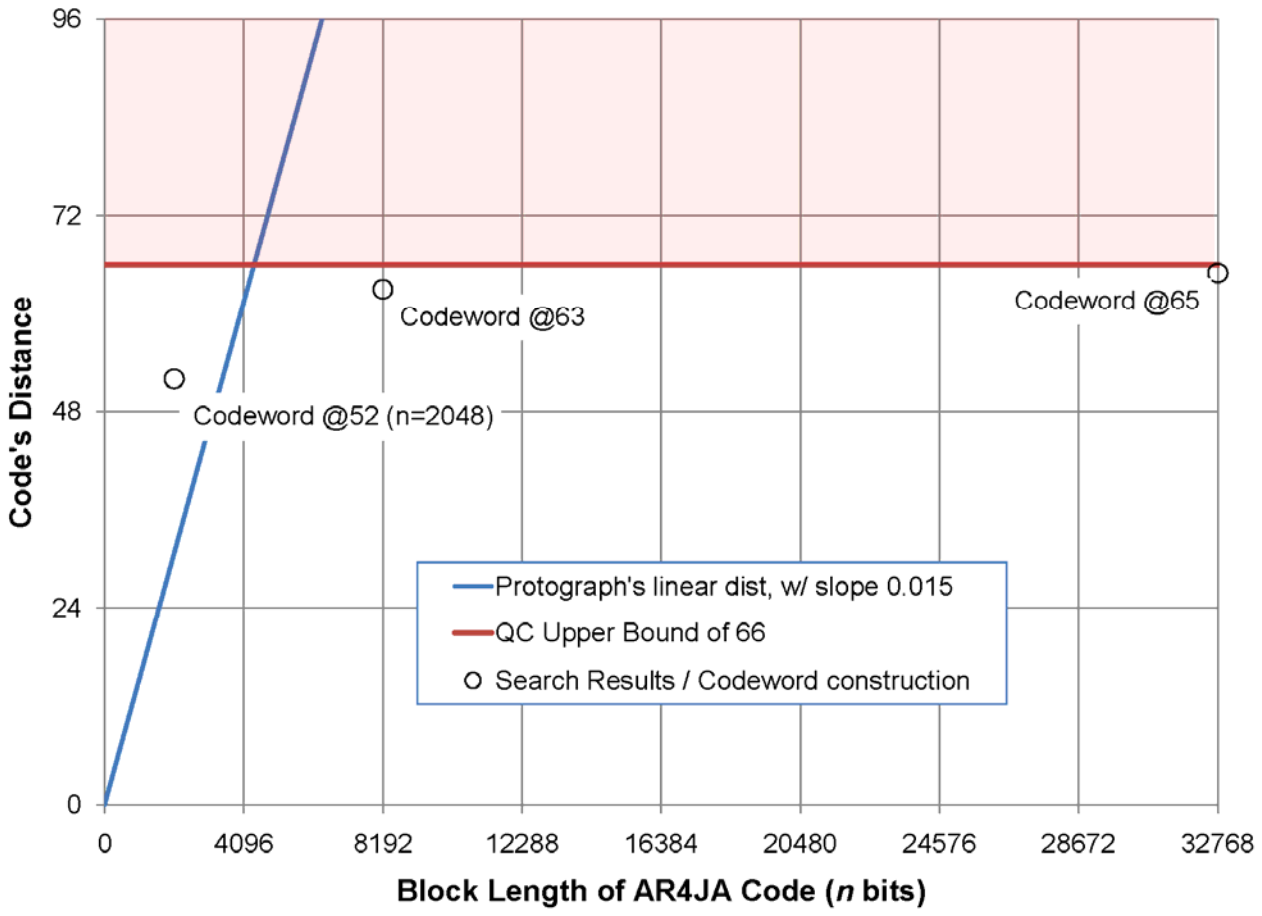

Figure 3. Distance vs. Block Length for rate 1/2 AR4JA.

The bounds developed here and [3] are useful tools in validating future QC LDPC code designs both punctured and un-punctured.

## VIII. Acknowledgment

The authors gratefully acknowledge contributions to the proofs by Dr. Pascal Vontobel.